%% file: tmi.tex
\documentclass[journal,twoside,web]{ieeecolor}
\pdfoutput=1
\usepackage{tmi}
\usepackage{cite}
\usepackage{amsmath,amssymb,amsfonts}
\usepackage{mathrsfs}
\usepackage{algorithm,algpseudocode}
\usepackage{eucal}
\usepackage{bbm}
\usepackage{tabularx,booktabs}
\newcolumntype{P}[1]{>{\centering\arraybackslash}p{#1}}
\newcolumntype{M}[1]{>{\centering\arraybackslash}m{#1}}

\usepackage{bm}
\usepackage{multicol}
\usepackage{multirow}
\usepackage{subfigure}

\usepackage{colortbl}
\usepackage{arydshln}
\usepackage{bbding}
\usepackage{graphicx}
\usepackage{textcomp}
\usepackage{algorithmicx}
\usepackage{algpseudocode}
\usepackage{float}

\newcommand{\marwan}[1]{{\color[rgb]{0,0,0}{}}}

\newcommand{\majorrevise}[1]{{\color[rgb]{0,0,0}{#1}}}

\newcommand{\etal}[1]{\emph{et al}.}

\makeatletter
\let\NAT@parse\undefined
\makeatother

\makeatletter
\let\NAT@parse\undefined
\DeclareRobustCommand\onedot{\futurelet\@let@token\@onedot}
\def\@onedot{\ifx\@let@token.\else.\null\fi\xspace}

\newcolumntype{b}{>{\centering\arraybackslash}X}
\newcolumntype{s}{>{\centering\arraybackslash\hsize=.67\hsize}X}
\newcolumntype{d}{>{\centering\arraybackslash\hsize=.52\hsize}X}
\newcolumntype{e}{>{\centering\arraybackslash\hsize=.58\hsize}X}
\newcolumntype{f}{>{\centering\arraybackslash\hsize=.29\hsize}X}
\newcolumntype{h}{>{\centering\arraybackslash\hsize=.4\hsize}X}
\newcolumntype{g}{>{\centering\arraybackslash\hsize=.8\hsize}X}

\algnewcommand{\LineComment}[1]{\State \(\triangleright\) #1}
\def\BibTeX{{\rm B\kern-.05em{\sc i\kern-.025em b}\kern-.08em
    T\kern-.1667em\lower.7ex\hbox{E}\kern-.125emX}}
\usepackage[allcolors=blue
]{hyperref}%

\usepackage[capitalize]{cleveref}

\crefname{section}{Sec.}{Secs.}
\Crefname{section}{Section}{Sections}
\Crefname{table}{Table}{Tables}
\crefname{table}{Tab.}{Tabs.}

\begin{document}

\title{FoPro-KD: Fourier Prompted Effective Knowledge Distillation for Long-Tailed Medical Image Recognition}

\author{Marawan Elbatel, Robert Martí and  Xiaomeng Li
\thanks{
M. Elbatel is with the Department of Electronic and Computer Engineering, The Hong Kong University of Science and Technology, and also with the Computer Vision and Robotics Institute, University of Girona. 
R, Martí is with the Computer Vision and Robotics Institute, University of Girona. 
X. Li is with the Department of Electronic and Computer Engineering at the Hong Kong University of Science and Technology, Hong Kong SAR, China and is also with HKUST Shenzhen-Hong Kong Collaborative Innovation Research Institute, Futian, Shenzhen.
M.E is partially funded by the EACEA Erasmus Mundus grant. RM is partially funded by the research project PID2021-123390OB-C21 funded by the Spanish Science and Innovation Ministry. This work is conducted with joint supervision from R.Martí and X.Li.
This work is partially supported by grants from Foshan HKUST Projects (Grants FSUST21-HKUST10E and FSUST21-HKUST11E) and the Project of Hetao Shenzhen-Hong Kong Science and Technology Innovation Cooperation Zone (HZQB-KCZYB-2020083).  Corresponding author: Xiaomeng Li (email: eexmli@ust.hk). 
}
\thanks{Copyright (c) 2021 IEEE. Personal use of this material is permitted. Permission from IEEE must be obtained for all other uses, including reprinting/republishing this material for advertising or promotional purposes, collecting new collected works for resale or redistribution to servers or lists, or reuse of any copyrighted component of this work in other works.}}

\maketitle
 
\begin{abstract}
Representational transfer from publicly available models is a promising technique for improving medical image classification, especially in long-tailed datasets with rare diseases. However, existing methods often overlook the frequency-dependent behavior of these models, thereby limiting their effectiveness in transferring representations and generalizations to rare diseases. In this paper, we propose FoPro-KD, a novel framework that leverages the power of frequency patterns learned from frozen pre-trained models to enhance their transferability and compression, presenting a few unique insights: 1) We demonstrate that leveraging representations from publicly available pre-trained models can substantially improve performance, specifically for rare classes, even when utilizing representations from a smaller pre-trained model. 2) We observe that pre-trained models exhibit frequency preferences, which we explore using our proposed Fourier Prompt Generator (FPG), allowing us to manipulate specific frequencies in the input image, enhancing the discriminative representational transfer. 3) By amplifying or diminishing these frequencies in the input image, we enable Effective Knowledge Distillation (EKD). EKD facilitates the transfer of knowledge from pre-trained models to smaller models. Through extensive experiments in long-tailed gastrointestinal image recognition and skin lesion classification, where rare diseases are prevalent, our FoPro-KD framework outperforms existing methods, enabling more accessible medical models for rare disease classification. Code is available at \url{https://github.com/xmed-lab/FoPro-KD}.

\end{abstract}

\begin{IEEEkeywords}
Visual Prompting, Knowledge Distillation, Long Tailed Learning.
\end{IEEEkeywords}

\input{tmi-1-Introduction}
\input{tmi-2-RelatedWorks}
\input{tmi-3-Method}

\input{tmi-4-Experiment}

\input{tmi-5-Ablation}

\input{tmi-6-DiscussionAndConc}

\bibliographystyle{ieeetr}
\small\bibliography{tmi}
\clearpage
\clearpage

\end{document}

%% file: tmi-1-Introduction.tex
\section{Introduction}
\label{sec:introduction}
\begin{figure}[t]
\centering 
\includegraphics{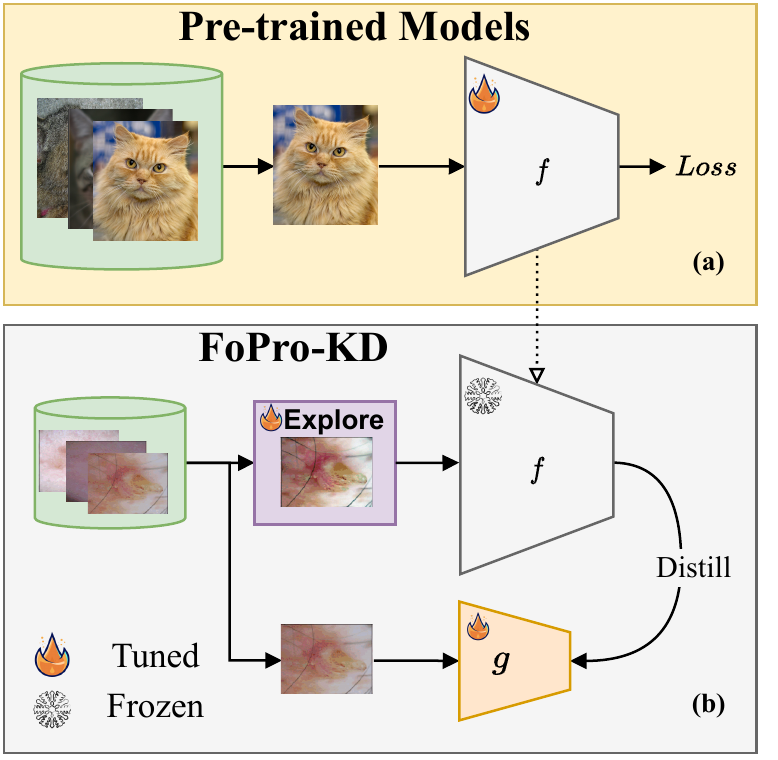}
\caption{(a) The pre-trained model  ``free lunch'' assumes specific frequency patterns in input data.
(b) Our FoPro-KD approach explicitly queries the model to identify meaningful frequency patterns for distillation. 
}
\label{fig:cover}
\end{figure}

\IEEEPARstart{C}{onvolutional}
 neural networks (CNNs) have shown remarkable performance in medical image classification. However, the scarcity of labeled medical image datasets can limit their applicability, particularly in datasets with long-tailed distributions where rare classes are present. Transfer learning has emerged as a promising approach to
 tackle this challenge by fine-tuning pre-trained models on natural images to medical image datasets. 
 Nevertheless, a crucial issue in transfer learning is to devise an efficient technique that preserves the generalization capabilities of large pre-trained models while remaining compact enough for practical deployment in a clinical environment.

Publicly available pre-trained models, such as CLIP~\cite{Radford2021LearningTV_clip}, MoCo~\cite{He2020MomentumCF_moco}, and BYOL~\cite{BYOL}, have attracted considerable attention in the medical imaging community due to their promising generalization capabilities. However, the extensive model complexity and significant computational resource requirements associated with these pre-trained models can limit their applicability in clinical settings in low infrastructure, point-of-care testing, and edge devices.
Moreover, fine-tuning (FT) these models on smaller, long-tailed medical image datasets offers reduced performance on the tail classes / rare diseases compared to linear probing (LP); see comparisons in Table~\ref{tbl:abl:ekd_small}. 
Therefore, it is highly demanded to develop an effective transfer learning approach to leverage the generalization capabilities of large pre-trained models while maintaining performance for tail classes in the target datasets. \majorrevise{In this paper, we use the term ``free lunch models'' to describe publicly accessible models pre-trained on natural images.}

\majorrevise{\majorrevise{``Free lunch models''} possess inherent traits associated with their preferred input frequencies and semantics, which might not harmonize effectively with the characteristics of the target datasets, leading to suboptimal performance when these preferences are not met. Recently, Yu~\etal~\cite{yu2023tuning}, quantified the frequency bias in neural networks and proposed a method for guiding the network to tune its frequency by utilizing a Sobolev norm that expands the L2 norm. However, the specific frequency patterns captured by \majorrevise{``free lunch models''} during the pre-training stage remained unexplored.

In this paper, we aim to explore the frequency-dependent behavior of \majorrevise{``free lunch models''} to manipulate the input data to enhance or diminish them, optimizing the transfer of representations from a ``free lunch model'', $f$, on natural images to a smaller target model, $g$, as illustrated in~\Cref{fig:cover}. This allows us to enhance the performance of the target model in downstream long-tailed medical applications, such as gastrointestinal image recognition and skin lesion classification.}

To this end, we propose a novel method called FoPro-KD (Fourier Prompted Effective Knowledge Distillation) to enhance the transferability of \majorrevise{``free lunch models''} from natural images to long-tailed medical image classification tasks. FoPro-KD consists of two stages: exploration and exploitation, which aims to find the frequency patterns that best suit the \majorrevise{``free lunch models''} and exploit this information to downstream tasks. In the exploration stage, we introduce a Fourier prompt generator (FPG) to unleash the frequency patterns based on the frozen pre-trained model, conditional on the target medical domain, for effective representational transfer. In the exploitation stage, the FPG generates targeted perturbations as Fourier amplitude spectral prompts for effective knowledge distillation (EKD).
EKD compresses the generalization capabilities of ``free lunch models'' into smaller medical imaging models more efficiently, enhancing tail classes / rare diseases recognition. To promote diverse patterns, we utilize adversarial knowledge distillation (AKD) to facilitate the exploration and exploitation process by iteratively learning the FPG.



Our proposed method represents a novel approach in the field of medical imaging in generating targeted perturbation to \majorrevise{``free lunch models''}. Rather than synthesizing worst-case images as in the literature of adversarial domain adaptation, we utilize a source-free frozen pre-trained model trained on natural images to learn Fourier spectrum amplitudes that are necessary for exploring these free lunch models. By exploring the frequency patterns learned by the pre-trained model, which we find to be efficient in exploiting its generalization capabilities, we can leverage free lunch models' generalization capabilities for a target medical imaging dataset without fine-tuning these models on the target dataset.

The main contributions of this work can be summarized as the following:
\begin{itemize}

\item  We demonstrate that effective knowledge distillation (EKD) from frozen pre-trained models on natural images to a target smaller medical imaging model can be just as effective as traditional long-tailed methods, thanks to their generalization capabilities.

\item We show that our generated Fourier prompts are highly effective in generating targeted perturbations that can further improve the generalization capabilities of our proposed EKD, particularly in long-tailed medical image classification tasks.

\item We introduce a novel framework called FoPro-KD, which achieves state-of-the-art performance on two long-tailed medical image classification benchmarks, demonstrating the effectiveness of our method in improving the transferability of pre-trained models to medical imaging tasks.
\end{itemize}

\begin{figure*}[ht]
\centering
\includegraphics[width=\textwidth]{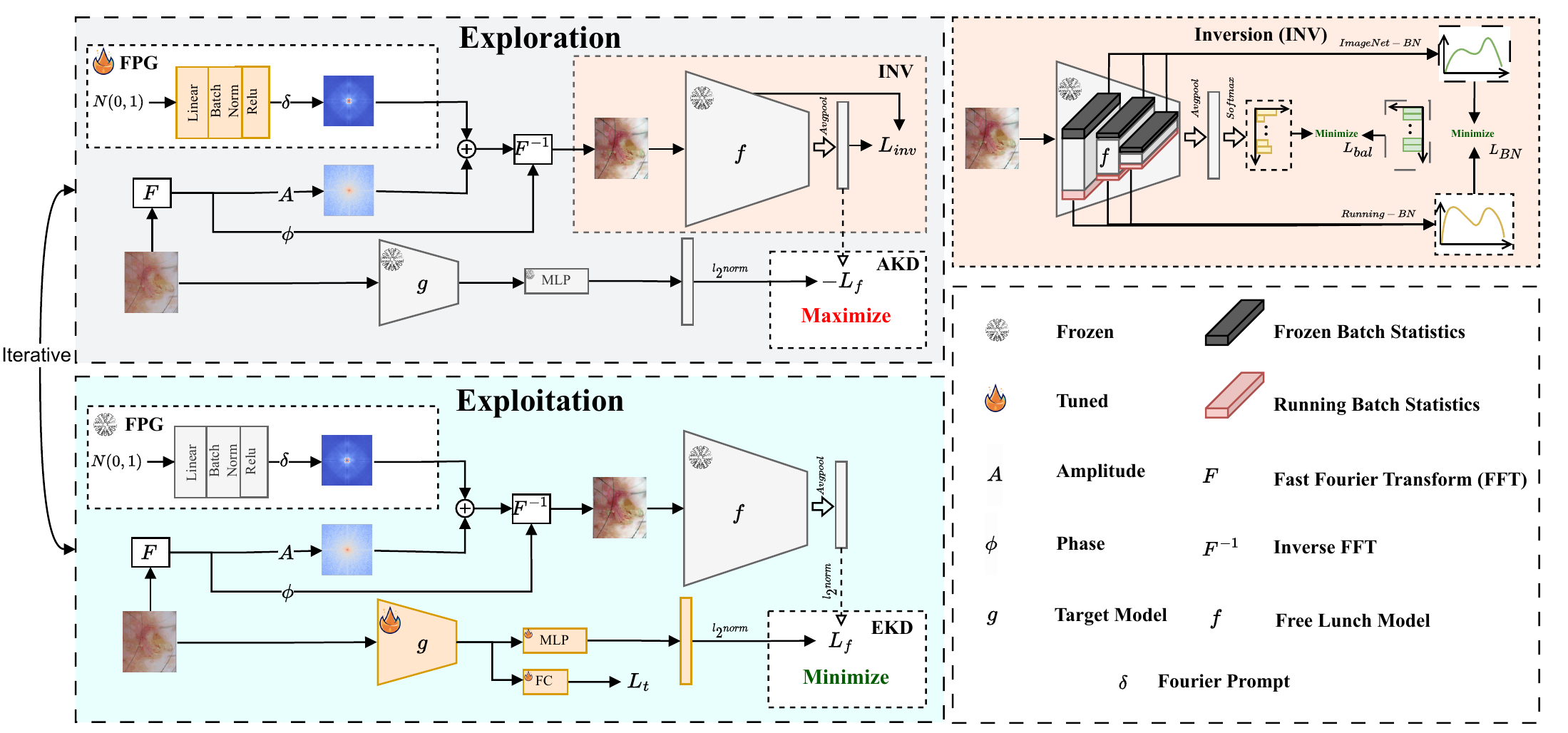}
\caption{
Our proposed FoPro-KD framework has two phases: exploration and exploitation. In the exploration phase, the FPG generates Fourier prompts to capture frequency patterns of the frozen pre-trained model $f$. In the exploitation phase, the proposed effective knowledge distillation (EKD) module distills the knowledge from $f$ into the target model $g$, guided by the Fourier prompt generator (FPG). Our framework can iteratively alternate between the exploration and exploitation phases using adversarial knowledge distillation (AKD) to enhance representation distillation and learning efficiency of $g$.}

\label{fig:our_method}
\end{figure*}


%% file: tmi-2-RelatedWorks.tex
\section{Related Work}
In this section, we review the literature related to transfer learning with prompt tuning, adversarial domain adaptation, and long-tailed learning methods. We highlight the relevant works in these areas and discuss their contributions.
\subsection{Transfer Learning}

In recent years, transfer learning and fine-tuning have been extensively studied in the literature, focusing on adapting the feature extractor to fit the target task. However, such approaches can deviate from pre-trained features, resulting in a trade-off between the performance of the majority class (in-distribution or IID) and the rare class (out-of-distribution or OOD). To mitigate similar tradeoffs on IID and OOD datasets, Kumar~\etal~\cite{kumar2022finetuning_LP_FT} proposed a simple variant of initializing the head with a linear probed version followed by full fine-tuning. Nevertheless, these methods can suffer from deviating semantics and extreme overfitting on long-tailed problems when fully fine-tuning \majorrevise{``free lunch models''}. Prompt tuning arises in vision to address these issues for efficiently fine-tuning large models in vision tasks, similar to natural language processing (NLP). Jia~\etal~\cite{jia2022vpt} proposed Vision Prompt \majorrevise{Tuning} (VPT), which adds prompts to vision transformers and exploits the transformer's location-invariant features for effective fine-tuning. Similar to NLP prompt tuning, Dong~\etal{}~\cite{dong2023lpt} explored the use of prompt learning for the effective transfer of pre-trained vision transformers for long-tail natural image classification. 

These methods are specially tailored to vision transforms similar to NLP, failing to find an efficient prompt for transforming the knowledge of CNN vision-pre-trained models, which are important for medical imaging classification. Recently, Bai~\etal~\cite{bai2022improving_vision_freq} found that a CNN teacher can benefit vision transformers to fit high-frequency components and proposed \majorrevise{high-frequency adversarial training for vision transformers,} to augment images' high-frequency components towards improving vision transformers generalization capabilities. 
Prompt tuning for CNN models can be related to the literature on adversarial learning and domain adaptation.

\subsection{Adversarial learning}
Adversarial learning has emerged as a popular approach for domain adaptation (DA) and domain generalization (DG).  To achieve DA, Huang~\etal~\cite{huang2021rda} proposed a method that generates adversarial examples from the source dataset and fine-tunes the model on the target dataset using both adversarial and clean examples. Similarly, Kim~\etal~\cite{kim2023domain} modeled DG as DA to adversarially generate worst-case targets from the source dataset. Chen~\etal~\cite{Chen2022MaxStyle} proposed MaxStyle as an adversarial realistic data augmentation utilizing an auxiliary image decoder for robust medical image segmentation. For source-free unsupervised domain adaptation (SFUDA), Hu~\etal~\cite{hu2022prosfda} proposed to learn a domain-aware prompt adversarially for a UNet-based model. More recently, Wang~\etal~\cite{Wang2023FVPFV}, inspired by Fourier style mining~\cite{YANG2022102457_fourier_mining}, proposed to learn a low-frequency Fourier visual prompt for SFUDA that excelled in segmentation performance. However, all these methods are restricted to source and target datasets trained for the same closed-set task and often rely on increasing noise to synthesize adversarial examples in DG or on bridging the gap between datasets in DA. Their approaches do not explicitly leverage the frequency patterns captured by \majorrevise{``free lunch models''} on natural images during their pre-training stage, which \majorrevise{could} aid in representational learning, especially for long-tailed datasets

\subsection{Long-Tail Learning}
Long-tailed distributions, characterized by severe class imbalance where minority classes are significantly outnumbered by majority classes, are common in many medical imaging tasks, such as skin-lesion classification and gastrointestinal image recognition~\cite{Borgli2019HyperKvasirAC, Combalia2019BCN20000DL_isic2019,Tschandl2018TheHD_ham2018}. Such class imbalance poses challenges for training accurate models, and various approaches have been proposed to address this issue, including data augmentation techniques, re-sampling and re-weighting schemes, and curriculum-based methods. Data augmentation techniques aim to regularize the model by incorporating regularization techniques to enhance the model's representations. For example, Zhang~\etal~\cite{zhang2018mixup} proposed MixUp offering information augmentation to regularize training. However, such regularization needs to be coupled with a balancing scheme to account for the huge class imbalance. Galdran~\etal~\cite{10.1007/978-3-030-87240-3_31_balanced_mixup} proposed Balanced-Mixup, a simple variant of MixUp using class conditional sampling that has compelling capabilities for highly imbalanced medical image classification. Moreover, data augmentation methods usually need to be coupled with different loss re-weighting strategies to account for the label distribution shift that can arise over the test set. Class balancing loss (CB)~\cite{cui2019class_cb_loss},  Label distribution margin (LDAM)~\cite{cao2019learning_LDAM}, and balanced-softmax (BSM)~\cite{Ren2020balms} was proposed as modified re-weighting strategies for training models for long-tailed learning. However, these methods often have limitations, such as not effectively addressing the extreme bias from \majorrevise{the majority} classes. To address such bias, Kang~\etal~\cite{Kang2020Decoupling} found that the classifier is the major bottleneck for the \majorrevise{majority} classes bias in long-tail learning and proposed a two-stage learning approach that decouples the feature extractor representations from the classifier through a plug-in classifier re-training (cRT). Although cRT increased the performance of multiple long-tailed methods, it did not solve the intra-class imbalance that can restrict the representation extraction~\cite{Zhao2021WellclassifiedEA_decouple_problem}. To address the intra-class imbalance, Tang~\etal~\cite{10.1007/978-3-031-20053-3_41_IF_dual} proposed invariant feature learning (IFL) through dual environment learning and re-sampling techniques. On the other hand, methods based on curriculum learning, requiring a pre-training stage on the target dataset to extract meaningful representation followed by utilizing these representations, have achieved state-of-the-art performance for long-tailed learning.  \majorrevise{Curriculum-based methods~\cite{10.1007/978-3-031-16437-8_44_monash_sampling_LT,ZHANG202336_bkd} rely on increasing the complexity of the task gradually, adopting a progressive approach. Typically, these methods employ a two-stage training framework, where a representation is extracted over the target dataset in an initial step, followed by utilizing this representation in another step. For instance, Ju~\etal~\cite{10.1007/978-3-031-16437-8_44_monash_sampling_LT} trains a self-supervised teacher model over the target dataset in the initial stage, followed by utilizing the teacher’s representations for difficulty-aware sampling for each class. Zhang~\etal~\cite{ZHANG202336_bkd} achieved SOTA in multiple long-tailed datasets by a two-stage framework.} First, by pre-training a teacher model on the target dataset to capture the target dataset representations, followed by a balanced knowledge distillation (BKD) to guide a student model. \majorrevise{The motivation behind~\cite{ZHANG202336_bkd} stems from the role played by the target dataset in acquiring representations that exhibit generalizability. Additionally, Zhang~\etal~\cite{ZHANG202336_bkd} incorporates class priors during training of the student model, facilitating the learning of tail classes. Despite the impressive performance of \majorrevise{``free lunch models''} known for their generalizable representations on natural images, the aforementioned methods have not leveraged these capabilities. Instead, they primarily focus on a narrow knowledge extraction basis from the target dataset, without fully harnessing the benefits of pre-training and the rich knowledge derived from natural images. Notably, pre-training on natural images has demonstrated compelling performance in medical imaging~\cite{ding2022free}.
}


In our work, we re-visit long-tailed learning in medical imaging from a free lunch perspective. We demonstrate that the generalizable features from publicly available pre-trained models on natural images can be comparable to different long-tail methods without additional pre-training or fine-tuning of these free lunch models. In addition, we find that these free lunch models have a preferred frequency basis (i.e. styles) for their input that can restrict their distillation in many tasks. To address such preferred styles, we propose to explore these preferred styles through effective prompting on a frequency basis. By exploring the \majorrevise{``free lunch models''} frequency patterns and iteratively distilling such knowledge, we can recycle and compress these \majorrevise{``free lunch models''} with no additional training to the target medical task, our approach can be easily utilized with different long-tailed learning schemes as a free lunch distillation, achieving SOTA on multiple long-tailed medical imaging datasets.

%% file: tmi-3-Method.tex
\section{Method}

\Cref{fig:our_method} shows the framework for our proposed FoPro-KD. The training of FoPro-KD consists of two stages: an exploration stage and an exploitation stage. In the exploration stage, we train one linear layer as a Fourier Prompt Generator (FPG) to generate Fourier amplitude spectral prompts, $\delta$, conditional on our target medical data, allowing us to explore the representations of the free lunch model, $f$, by explicitly asking what frequency patterns on the input lead to meaningful representations. This is done while freezing $f$, pre-trained on a natural imaging dataset (ex: MoCov2 on ImageNet~\cite{He2020MomentumCF_moco}). In the exploitation stage, we effectively distill these generalizable representations to a smaller target medical imaging model, $g$ through our proposed Effective Knowledge Distillation (EKD). To make the Fourier prompts more diverse while being representative of $f$, we perform multiple iterations of the exploration and exploitation stages by an Adversarial Knowledge Distillation (AKD). This allows us to effectively exploit the generalization capabilities of \majorrevise{``free lunch models''} and compress them into smaller student networks that are useful for practical medical imaging deployment in a clinical setting. Our framework provides a scalable and efficient approach to distilling knowledge from \majorrevise{``free lunch models''}, with potential applications in various medical imaging tasks.

\subsection{Fourier Prompt Generation}

\begin{figure}
    \centering
\includegraphics[width=\columnwidth]{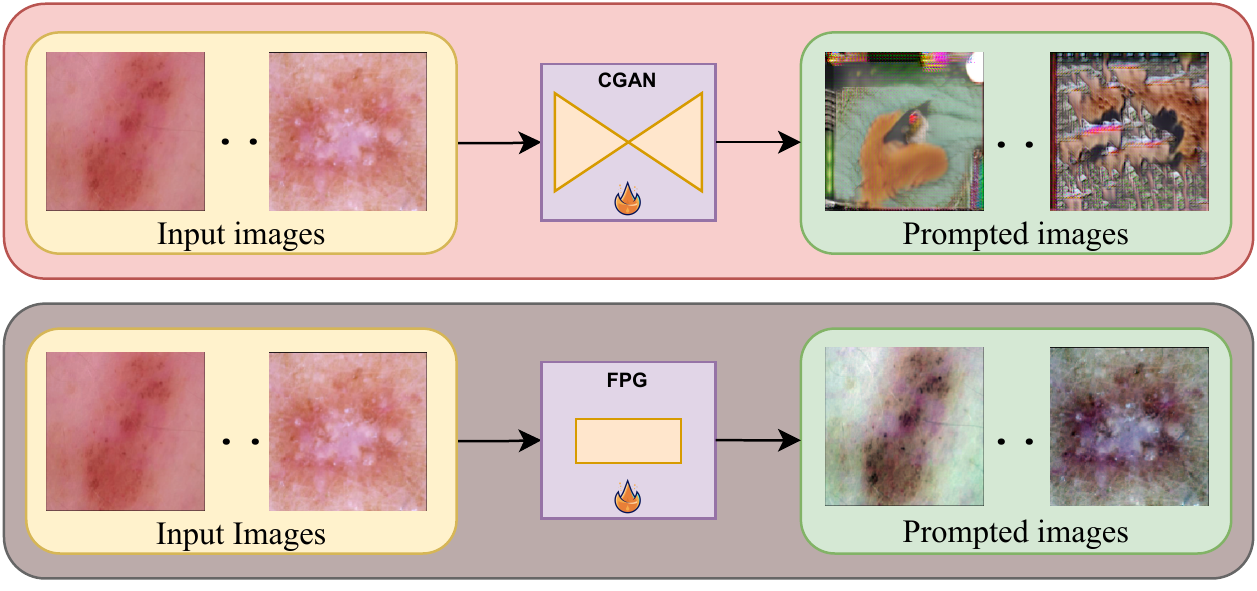}
    \caption{Using a conditional GAN (CGAN) to manipulate the input dataset changes the image semantics in highly informative regions compared to surpassing or amplifying certain frequencies in these regions with Fourier Prompt Generator (FPG).}
    \label{fig:cgan_vs_fopro}
\end{figure}

To attain optimal representational transfer, \majorrevise{``free lunch models''} necessitate input data that closely align with their preferences. In this regard, training a conditional generative adversarial network (CGAN)~\cite{mirza2014conditional} to guide the target dataset towards these preferences can substantially modify the semantics of the dataset. As shown in~\Cref{fig:cgan_vs_fopro}, training a CGAN with deep inversion causes modification in the semantics of the target dataset in the highly informative regions conditional on the semantics of the pre-training dataset, ImageNet~\cite{deng2009imagenet}.

Recent research by Yu~\etal~\cite{yu2023tuning} has shown theoretically that neural networks can be sensitive to certain frequencies without explicitly considering the frequency patterns captured during pre-training deep neural networks (DNNs). Therefore, we aim to explore this frequency-dependent behavior of CNNs and enable \majorrevise{``free lunch models''} to output representations through prompting on a frequency basis, which is facilitated by our proposed Fourier Prompt Generator (FPG).

FPG employs a random noise vector, $z$, to generate a three-dimensional Fourier amplitude prompts, $\delta=FPG(z)$, one for each channel respectively, enabling the modification of the target dataset by emphasizing or suppressing specific frequency patterns preferred and captured by ``free lunch models'' on the source natural images dataset. Although these preferred patterns relied on the deep learning dynamics  of the ``free lunch models'', the FPG can be trained to unleash such patterns and generate Fourier prompts that are the preference of the ``free lunch model'' conditioned on our target medical dataset. This feature plays a critical role in effective knowledge distillation.

Let the Fourier decomposition of an image x be F(x), which consists of the amplitude A and phase $\phi$ components:

\begin{equation}
F(x) = A \odot e^{i\phi}
\end{equation}

To interpolate the Fourier amplitude between the input image and the generated Fourier prompt, we use a mixing coefficient, denoted by $\alpha$ and sampled uniformly from 0 to 1, resulting in a new Fourier amplitude spectrum $\hat{A}$:


\begin{equation}
\hat{A}_{ij} = \alpha A_{ij}  + (1-\alpha) \delta_{ij}
\end{equation}

\noindent where $A_{ij}$ represents the Fourier amplitude of the input image, $\delta_{ij}$ represents the the generated Fourier prompt, and $ij$ are the indices of the Fourier coefficients. 

The modified Fourier coefficients are then transformed back using the inverse Fourier transform to generate the modified image, denoted by $\hat{x}$,

\begin{equation}
\hat{x} = F^{-1}(\hat{A} \odot e^{i\phi})
\label{eq:modified_img}
\end{equation}
where $F^{-1}$ denotes the inverse Fourier transform.

We train the Fourier prompt generator in the exploration phase while freezing all other modules. Specifically, we feed $\hat{x}$ to the \majorrevise{``free lunch model''}, $f$, and utilize the batch regularization technique that was first introduced in~\cite{9157759_loss_bn}. This technique minimizes the divergence between the feature statistics, which include the mean and variance of the features, and the corresponding batch normalization statistics by assuming a Gaussian distribution:
\begin{equation}
\mathcal{L}_{\text{BN}}(x) = \sum_{l \in f} D\Big(N\Big(\mu_l(\hat{x}), \sigma_l^2(\hat{x})\Big) \Big| N\Big(\mu_l, \sigma_l^2\Big)\Big),
\label{eq:bn_reg}
\end{equation}

\noindent where $D$ is the L2 divergence loss, $N(\mu_l(\hat{x}), \sigma_l^2(\hat{x}))$ is the feature statistics of the modified input batch $\hat{x}$, $N(\mu_l, \sigma_l^2)$ is the batch normalization statistics of the frozen model, $f$, and $l$ indexes the layers of $f$.

To better capture the \majorrevise{``free lunch model''} learned frequency patterns and avoid skewing in the learning of the Fourier Prompt Generator (FPG), we propose a regularization approach that encourages the synthesis of Fourier prompts with a more balanced distribution of activations across the final pre-classification features. This is achieved by maximizing the entropy of the free lunch model output towards a uniform distribution where each feature has an equal probability of being activated as 

\begin{equation}
\mathcal{L}_{\text{bal}} = \sum_{i=1}^{C} p_i \log p_i
\end{equation}

\noindent where $C$ is the dimension of the final pre-classification features, and $p_i$ is the $i$-th element of the softmax output $p$ of the \majorrevise{``free lunch model''} on the target modified data~$\hat{x}$. This approach avoids bias towards any particular feature and promotes the generalization ability of the learned Fourier prompts.

The final inversion loss $\mathcal{L}_{inv}$ to train the FPG module is defined as the combination of the batch normalization loss, $\mathcal{L}_{BN}$), and the balancing loss, $\mathcal{L}_{bal}$, as
\begin{equation}
\mathcal{L}_{inv} = \mathcal{L}_{BN} + \mu \ \mathcal{L}_{bal}
\label{eq:inversion_loss}
\end{equation}
where $\mu$ is the weighting factor for the balancing regularization. 

By combining this balanced regularization term with the batch statistics losses, the generated Fourier prompts can exhibit higher entropy while being specific to the \majorrevise{``free lunch model''} desired frequencies to better benefit the knowledge distillation. To ensure that the FPG-generated Fourier prompts produce valid Fourier amplitudes, we apply a Hermitian constraint.

The exploration phase ensures that the Fourier generator produces styles that are consistent with the preferred frequency patterns of the free lunch model while also avoiding overfitting specific styles.


Our training approach for the FPG can be seen as a deep inversion method in the literature of data-free knowledge distillation~\cite{fang2021contrastive_cmi}. However, our method is unique in the learnable and target objectives, in addition, conditioned on a cross-task target dataset, which makes it more challenging.

\subsection{Exploitation with Effective Knowledge Distillation}

\majorrevise{``Free lunch models''} available to the public possess remarkable generalization capabilities that can assist in the classification of rare diseases. It has been observed that performing linear probing on these models yields high-accuracy results on out-of-distribution (OOD) datasets. However, complete fine-tuning of these models may lead to distortion of these highly generalizable representations~\cite {kumar2022finetuning_LP_FT}. To this end, we propose Effective Knowledge Distillation (EKD), which aims to compress the generalization capabilities of the free lunch models while maintaining generalizable performance on the target data using a smaller model.

To achieve this, we utilize a small target model with a feature extractor $g(\cdot)$ to be trained on the target medical dataset, along with a \majorrevise{``free lunch model''}, $f(\cdot)$. To compare the latent features of the target model with those of the free lunch model, we add a 2-layer MLP on top of the smaller target feature extractor, $g(\cdot)$.

To generate the necessary encodings for distillation, we sample an image $x$ uniformly from the target dataset $\mathcal{D}$ and use prompt mixing following~\Cref{eq:modified_img} to obtain $\hat{x}$, while freezing FPG. This allows us to navigate the representation of $f$ based on the styles and frequencies it was trained on. 

From here, we generate two encodings: a \emph{ projection} $y = MLP(g(x))$ and a \emph{target representation} $t = f(\hat{x})$ from our target network and the \majorrevise{``free lunch model''}, respectively. We then L2-normalize both encodings and distill the information from the \majorrevise{``free lunch model''} to the smaller target model using a mean squared error loss as our distillation loss between both encodings, as 

\begin{equation}
    \mathcal{L}_{f} = 2 - 2\cdot{\langle y, t \rangle}
    \cdot
    \label{eq:cosine-loss}
\end{equation}

While previous approaches~\cite{chen2022simkd} \majorrevise{aim} to reduce the performance gap between the teacher and student models on the same task, our proposed distillation loss is designed to narrow the generalization capabilities of free lunch models to a different task, which our student model is being trained on. This approach can act as an implicit regularization technique, leveraging the discriminative generalization capabilities of \majorrevise{``free lunch models''} features for the tail classes. Specifically, our approach encourages the $g(\cdot)$ to generalize well to the tail classes of the target task, which may be rare and difficult to identify without additional guidance. We denote the exploitation loss, $\mathcal{L}_{exploit}$, to minimize at each training step as:
\begin{equation}
\mathcal{L}_{exploit} = \mathcal{L}_{t}+ \lambda_{f} ~\mathcal{L}_{f},
\label{eq:loss_total}
\end{equation}
where $\mathcal{L}_{t}$ is a target loss for long-tail learning~\cite{Ren2020balms}.

\subsection{Exploration with Adversarial Distillation}
To ensure that the learned Fourier amplitudes become more diverse while being representative of the natural image styles, thus alleviating any representational mode collapse issue in distillation, we propose to further enhance the Fourier prompt generation by navigating the latent space of the free lunch model with an iterative adversarial loss.

To achieve this, we propose maximizing the proposed effective knowledge distillation (EKD) loss between the free lunch model and the target model for iterative exploration. The final exploration loss, $\mathcal{L}_{explore}$, to be optimized is given by:

\begin{equation}
\mathcal{L}_{explore} = - \gamma \lambda{f}  \mathcal{L}_{f} + \mathcal{L}_{inv} \label{eq:adv_loss}
\end{equation}

Here we maximize the similarity between the free-lunch model and the target model, as described in~\Cref{eq:cosine-loss}. This adversarial loss is weighted by a hyperparameter $\gamma$, which determines the strength of the adversarial training. Unlike standard adversarial training, we aim to explore the free-lunch model, so we set $\gamma$ between 0 and 1, with an upper bound of the exploitation distillation factor $\lambda_f$. This is similar to the training of generative adversarial networks (GANs)~\cite{Goodfellow2014GenerativeAN}. To this end, we choose a value of $\gamma=0.3$ and provide an ablation study to validate our choice. $\mathcal{L}_{inv}$ ensures that the prompts generated by the Fourier prompt generator accurately represent the \majorrevise{``free lunch model''}.

%% file: tmi-4-Experiment.tex
\section{Experiments}

\subsection{Datasets}
\noindent \textbf{ISIC-LT} is a challenging long-tailed skin lesion classification dermatology dataset from ISIC~\cite{Combalia2019BCN20000DL_isic2019}. The dataset consists of eight classes and we create a long-tailed version of it following~\cite{10.1007/978-3-031-16437-8_44_monash_sampling_LT} using a Pareto distribution sampling approach. To ensure class imbalance and rare disease diagnosis, we set the class imbalance ratio to be {100, 200, 500} and select 50 and 100 images from each class for the validation and test sets respectively, from the remaining samples. \majorrevise{The relevant statistics for the training dataset split are presented in~\Cref{tbl:isic_lt_statistics}.} We assess the model performance on the held-out test set. Results for each method are averaged over 5 runs, each with a different sampled train, validation, and held-out test set. To assess the model performance on the balanced test set, we follow previous guides~\cite{reinke2022metrics} to report the Mathew-correlation coefficient (``MCC''), accuracy (``Acc''), and f1-score.

\begin{figure}[h]

    \centering
\includegraphics[width=\columnwidth]{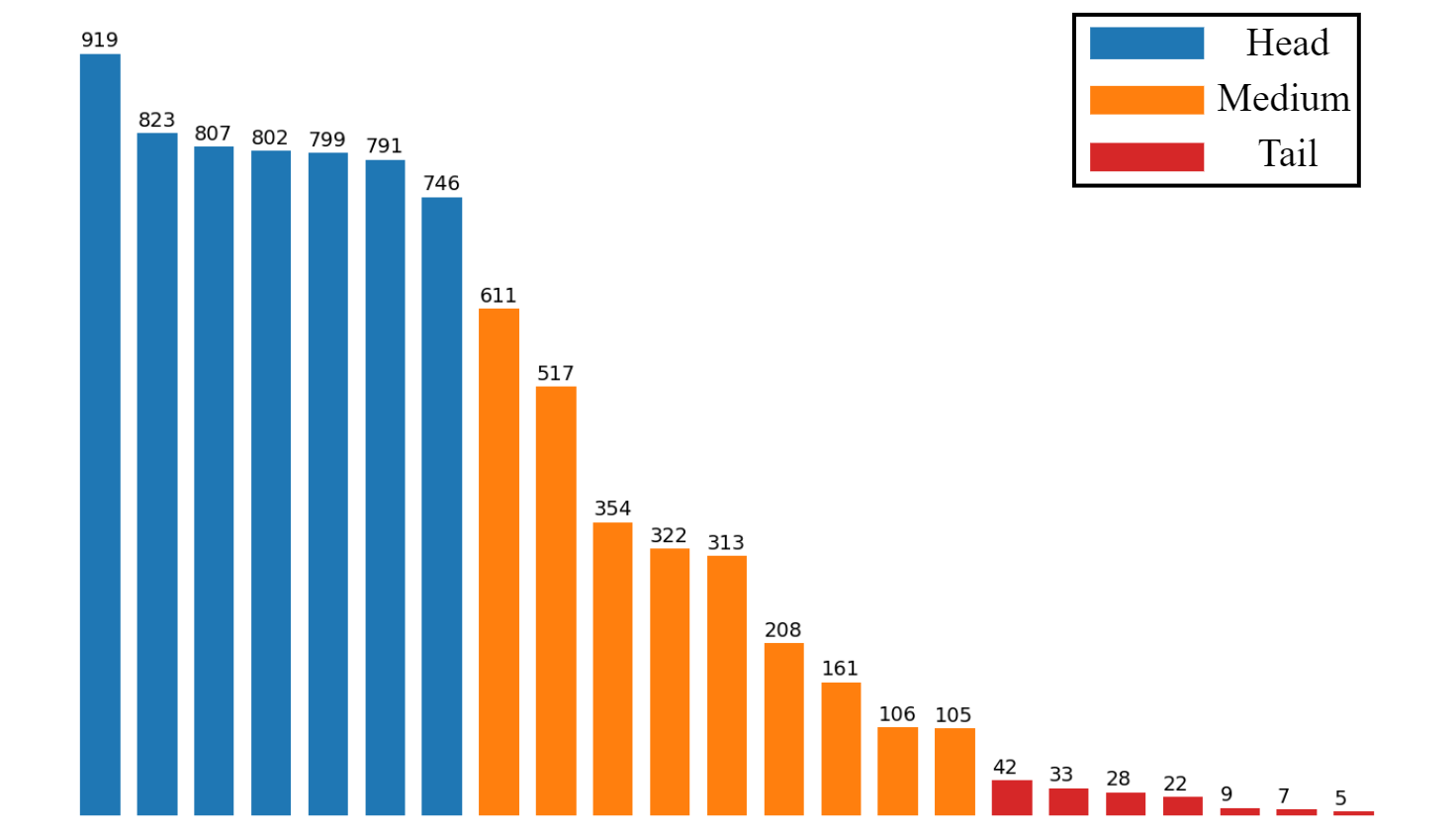}
    \caption{\majorrevise{The class distribution statistics of the gastrointestinal dataset for one training fold, spanning 23 classes. The classes are categorized into ``Head'', ``Medium'', and ``Tail'' for analysis purposes.}}
    \label{fig:gastro_data_stats}
\end{figure}

\noindent \textbf{Hyperkvasir} is a long-tailed dataset of 10,662 gastrointestinal tract images, consisting of 23 classes representing different anatomical and pathological landmarks and findings. To analyze the long-tailed distribution, we categorize the 23 classes into three groups: Head (with over 700 images per class), Medium (with 70 to 700 images per class), and Tail (with fewer than 70 images per class) based on their class counts~\majorrevise{as shown in~\Cref{fig:gastro_data_stats}}. Notably, the Tail class includes a distinct class for Barrett's esophagus, which presents as short segments and is considered a premalignant condition that may progress to cancer. Additionally, the Tail classes encompass two transitional grades of ulcerative colitis, an inflammatory bowel disease, and the terminal ileum, which confirms a complete colonoscopy but cannot be differentiated endoscopically from parts of the small bowel. Since the official test set only contains 12 classes, we follow the evaluation approach of BalMixUp~\cite{10.1007/978-3-030-87240-3_31_balanced_mixup} and assess our model's performance using a stratified 5-fold cross-validation method. To assess the performance with a high imbalance test set~\majorrevise{as most exisiting works~\cite{Zhang2021DeepLL_shot_based_division,10.1007/978-3-031-17027-0_3_shot_based_divison}}, we report the balanced accuracy ``B-Acc'' that considers the average per class accuracy and denote the performance of the few-shot division (``Head'', ``Medium'', ``Tail'') and their average results denoted as ``All''.

\begin{table}[h]
\centering
\majorrevise{
\caption{ISIC-LT Training Split. For all splits, the validation and test are 50 and 100 images from each class from the remaining samples. Full refers to the full dataset. Experiments are averaged over 5 runs, each with a different sampled train, validation, and held-out test set.}
    \resizebox{\columnwidth}{!}{
 \begin{tabular}{l|cccccccc}
        \toprule
    {\multirow{2}{*}{Split}} &\multicolumn{8}{c}{Class}  
    \\
    \cline{2-9}
        & NV&MEL&BCC&BKL&AK&SCC &VASC&DF\\
        \hline
        Full& 12,875&4,522 &  3,323& 2,624& 867& 628& 253& 239 \\
        \hline
        1:100&
        12,725&  4,372&  3,173&  1,788&   717& 478&   103& 89
        \\
        1:200& 12,725&4,372&2,833&1,329& 623& 292&   103&64
        
        \\
        1:500& 12,725&  4,372& 2,180&897&369&   152&62&25
        \\
        1:1000& 12,725&  4,372& 1,788&   666&   248&92&34&12
        \\
        1:2000& 12,725&4,346&1,467& 495&   167&56&19&6\\
        \bottomrule
    \end{tabular}
    \label{tbl:isic_lt_statistics}
    }}
\end{table}


\input{tmi-4-sota}

\subsection{Implementation Details}
For both datasets, we use checkpoints pre-trained on MoCo-RN50~\cite{He2020MomentumCF_moco} available online as the free lunch models trained on ImageNet for compressing its generalization capabilities unless otherwise stated. We use Adam optimizer with a learning rate of $3e-4$ for all methods on the ISIC-LT dataset. On the other hand, we follow~\cite{10.1007/978-3-030-87240-3_31_balanced_mixup} for the HyperKvasir dataset and use SGD with a cosine annealing scheduler~\cite{loshchilov2017sgdr_cosine} with a maximum learning rate of 0.01. For both datasets and all methods, we use a ResNet-18 as the target model with a batch size of 32 and apply augmentation techniques such as random crop and flipping. Images are resized to 224x224, and we train all methods until there is no further increase in the validation set \majorrevise{accuracy} for 20 epochs with a total of 100 epochs. To ensure a fair comparison between different methods, we keep all hyperparameters the same. We set $\lambda_{f}$ to 3, $\mu$ to 10, and $\gamma$ to 0.3 on both datasets. For every 5 training epochs exploited, we explore the \majorrevise{``free lunch model''} for one epoch to balance the training process.

\subsection{Baselines}
Our experimental evaluation compares the performance of our proposed FoPro-KD method against several state-of-the-art long-tailed learning approaches. Specifically, we evaluate (1) re-sampling (RS) and re-weighting (RW) techniques, (2) various data augmentation techniques including MixUp~\cite{zhang2018mixup}, and its balanced version (BalMixUp)~\cite{10.1007/978-3-030-87240-3_31_balanced_mixup}, specifically designed for medical image classification (3) Modified Loss re-weighting schemes including Class balancing (CB) loss~\cite{cui2019class_cb_loss}, and label-distribution-aware margin (LDAM) loss with curriculum delayed reweighting (DRW)~\cite{cao2019learning_LDAM}, and the balanced softmax (BSM)~\cite{Ren2020balms} (4) A recent curriculum-based method, balanced Knowledge Distillation (BKD) ~\cite{ZHANG202336_bkd}.

\subsection{Performance on ISIC-LT}
We present the performance of our proposed FoPro-KD approach for long-tailed skin lesion classification on the ISIC-LT dataset in~\Cref{tbl:sota_results_isic}. Our approach outperforms all baselines across all class imbalance ratios and evaluation metrics, demonstrating its effectiveness. FoPro-KD improves the performance of the naive cross entropy by 10.7\%, 12.4\%, and 12.4\% on the ``MCC'' over the balanced test set for class imbalance ratios of 1:100, 1:200, and 1:500, respectively. Compared to the baseline, BSM~\cite{Ren2020balms}, FoPro-KD improves the ``MCC'' being sensitive for class imbalance by 4.5\%, 5.6\%, and 3.7\% for imbalance ratios of 1:100, 1:200, and 1:500, respectively.  Furthermore, it increases the performance of the baseline, BSM~\cite{Ren2020balms}, by 3.7\%, 4.7\%, and 2.9\% on the ``Acc''  metric for class imbalance ratios of 1:100, 1:200, and 1:500, respectively. Compared to the best-performing baseline on imbalance ratios 1:200 and 1:500, BKD~\cite{ZHANG202336_bkd}, our method outperforms it by 6.0\%, 3.0\%, and 3.0\% on the ``MCC'' for the three imbalance ratios, respectively. Notably, our approach outperforms BKD on the f1-score with 3.6\% and 3.9\% performance gains over the imbalance ratios 1:200 and 1:500 without additional pre-training on the target dataset.

It is worth mentioning that our proposed EKD used with the naive cross-entropy loss \majorrevise{already} improves performance by 3.7\%, 3.9\%, and 4.3\% on the ``MCC'' metric for class imbalance ratios of 1:100, 1:200, and 1:500, respectively, without \majorrevise{using} FPG or special loss re-weighting or re-sampling, demonstrating the need to leverage the free lunch models for the long-tail problems in an effective way.

\subsection{Performance on HyperKvasir}
\majorrevise{
We present the experimental results of our long-tailed approach to gastrointestinal image recognition  in~\Cref{tbl:sota_results_gastro}. {It is worth noting that, unlike the skin-lesion dataset, the test set of the gastrointestinal dataset exhibits a significant class imbalance. Consequently, balanced accuracy (``B-Acc'') is reported as a robust metric for assessing performance adhered to the guidelines~\cite{reinke2022metrics}. Our method excels in this regard, outperforming all other methods by a considerable margin. Specifically, our approach, FoPro-KD, outperforms the naive cross-entropy (CE) method and the baseline, BSM~\cite{Ren2020balms}, by 5.8\% and 2.9\% respectively on the ``B-Acc'' metric. Furthermore, our FoPro-KD outperforms the state-of-the-art 
curriculum-based method BKD~\cite{ZHANG202336_bkd} on the HyperKvasir dataset by 3.8\% on the ``B-Acc''. BKD~\cite{ZHANG202336_bkd} relies on distilling a pre-trained teacher model over the target dataset, which can amplify bias over the head classes if the teacher model fails to capture the tail classes in the pre-training stage. In contrast, our approach presents a novel perspective by leveraging the discriminative generalizable features of free lunch models, achieving remarkable improvements.

Similar to existing work~\cite{Zhang2021DeepLL_shot_based_division,10.1007/978-3-031-17027-0_3_shot_based_divison}, we denote the average results of the three categories (``Head'', ``Medium'', ``Tail'') as ``All'' to demonstrate the performance of a model that consistently performs well across the three categories regardless of differences in sample sizes or class distributions within each group. It is crucial to understand that evaluating the results of the ``Head'', ``Medium'', and ``Tail'' classes independently in an imbalanced test set can lead to unreliable conclusions, being significantly influenced by the majority classes samples. On the ``All'' metric, our approach, FoPro-KD, outperforms the naive cross-entropy (CE) method, BSM~\cite{Ren2020balms}, and BKD~\cite{ZHANG202336_bkd} by 7.0\%, 2.9\%, and 4.3\% respectively, achieving the highest sensitivity for the tail classes (31.9\%) and highlighting its ability to capture rare diseases.
}



}

%% file: tmi-4-sota.tex
\definecolor{ao}{rgb}{0.01, 0.75, 0.24}

\begin{table*}[htbp]
  \centering
  \caption{Experimental results on long-tailed skin lesion classification (ISIC-LT) considering different class imbalance ratios. The methods used include naive cross-entropy (CE), class sampling (RS), and loss re-weighting (RW). The reported results are averaged over 5 runs on a balanced held-out test set.}
  \label{tbl:sota_results_isic}
  \resizebox{1\textwidth}{!}{
  \begin{tabular}{l|ccc|ccc|ccc}
    \toprule
    \multirow{3}{*}{Method} & \multicolumn{9}{c}{Class Imbalance Ratio} \\ \cline{2-10}
     & \multicolumn{3}{c|}{1:100} & \multicolumn{3}{c|}{1:200} & \multicolumn{3}{c|}{1:500} \\
     \cline{2-10}
     & MCC & Acc & F1-Score &
      MCC & Acc & F1-Score &
       MCC & Acc & F1-Score \\
    \midrule
    
CE &57.64 ($\pm$1.6)&62.15 ($\pm$1.4)&65.52 ($\pm$1.4)&53.71 ($\pm$1.7)&58.33 ($\pm$1.5)&62.72 ($\pm$1.2)&44.9 ($\pm$2.2)&50.22 ($\pm$1.9)&55.83 ($\pm$2.0) \\

RS &59.46 ($\pm$1.0)&63.9 ($\pm$0.9)&67.04 ($\pm$0.6)&55.53 ($\pm$1.6)&60.35 ($\pm$1.5)&63.71 ($\pm$1.4)&48.54 ($\pm$1.4)&53.73 ($\pm$1.2)&59.15 ($\pm$1.1)\\

RW  &56.03 ($\pm$2.3)&61.2 ($\pm$1.9)&63.17 ($\pm$2.2)&52.22 ($\pm$1.6)&57.95 ($\pm$1.4)&59.48 ($\pm$1.5)&46.77 ($\pm$0.4)&52.8 ($\pm$0.4)&55.36 ($\pm$0.7)\\

\textbf{EKD (ours)}&61.37 ($\pm$1.8)&65.42 ($\pm$1.6)&68.49 ($\pm$1.5)&57.57 ($\pm$1.1)&61.9 ($\pm$0.9)&65.41 ($\pm$1.2)&49.16 ($\pm$1.9)&54.2 ($\pm$1.8)&59.24 ($\pm$1.4)\\

\hline
CB~\cite{cui2019class_cb_loss} &57.28 ($\pm$2.3)&62.23 ($\pm$2.1)&64.36 ($\pm$1.6)&53.58 ($\pm$2.1)&58.9 ($\pm$2.1)&61.27 ($\pm$1.6)&47.16 ($\pm$1.2)&53.17 ($\pm$1.1)&55.9 ($\pm$1.6)\\

LDAM-DRW~\cite{cao2019learning_LDAM} &60.27 ($\pm$0.7)&64.88 ($\pm$0.6)&66.17 ($\pm$0.7)&55.85 ($\pm$1.6)&60.98 ($\pm$1.5)&62.25 ($\pm$1.3)&50.34 ($\pm$1.1)&55.98 ($\pm$0.8)&57.95 ($\pm$1.3)\\

BSM~\cite{Ren2020balms} &\underline{63.88} 
 ($\pm$1.9)&\underline{68.15} ($\pm$1.7)&\underline{69.25} ($\pm$1.6)&60.47 ($\pm$1.6)&65.12 ($\pm$1.4)&66.2 ($\pm$1.2)&53.61 ($\pm$1.1)&59.02 ($\pm$1.0)&60.27 ($\pm$0.9) 
\\

\hline
MixUp~\cite{zhang2018mixup}&55.53 ($\pm$1.8)&59.91 ($\pm$1.9)&64.33 ($\pm$1.0)&48.96 ($\pm$2.1)&53.59 ($\pm$2.2)&59.68 ($\pm$1.5)&43.03 ($\pm$1.6)&48.12 ($\pm$1.5)&54.36 ($\pm$1.1) \\


BalMixup~\cite{10.1007/978-3-030-87240-3_31_balanced_mixup}&61.35 ($\pm$1.8)&65.5 ($\pm$1.5)&68.46 ($\pm$1.5)&56.36 ($\pm$3.9)&61.0 ($\pm$3.5)&64.37 ($\pm$3.5)&50.26 ($\pm$1.1)&55.3 ($\pm$1.1)&60.29 ($\pm$0.7) \\

\hline
BKD~\cite{ZHANG202336_bkd}  &62.24 ($\pm$1.6)&66.55 ($\pm$1.6)&68.35 ($\pm$0.9)&\underline{63.06} ($\pm$1.4)&\underline{67.42} ($\pm$1.2)&\underline{68.32} ($\pm$1.3)&\underline{54.25} ($\pm$1.3)&\underline{59.59} ($\pm$1.1)&\underline{60.5} ($\pm$1.2)\\

\hline

\textbf{FoPro-KD (ours)}&\textbf{68.33 ($\pm$2.3)}&\textbf{71.8 ($\pm$2.0)}&\textbf{73.88 ($\pm$1.9)}&\textbf{66.08 ($\pm$1.5)}&\textbf{69.8 ($\pm$1.3)}&\textbf{71.91 ($\pm$1.2)}&\textbf{57.33 ($\pm$1.5)}&\textbf{61.9 ($\pm$1.5)}&\textbf{64.43 ($\pm$1.3)} \\

    \bottomrule
  \end{tabular}}
\end{table*}
\begin{table}[htbp]
  \centering
  \caption{Experimental results on long-tailed Gastrointestinal image recognition. The top-1 accuracy is reported using a shot-based division (``Head'', ``Medium'', ``Tail'') to address test set imbalance, and their average ``All'', along with the resilient metric ``B-Acc'' for class imbalance.}
  \label{tbl:sota_results_gastro}
  \resizebox{1\columnwidth}{!}{
  \begin{tabular}{l|ccc|cc}
    \toprule
    \multirow{2}{*}{Method}
     & \multicolumn{5}{c}{Metrics} 
     \\ \cline{2-6}
     & Head & Medium & Tail &All&B-Acc\\
    \midrule 

CE &93.14 ($\pm$0.7)&74.7 ($\pm$1.2)&4.05 ($\pm$4.8)&57.3 ($\pm$1.3)&58.81 ($\pm$1.1) \\

RS &88.89 ($\pm$3.9)&72.37 ($\pm$3.2)&11.38 ($\pm$10.4)&57.55 ($\pm$1.8)&58.84 ($\pm$1.6) \\

RW &87.43 ($\pm$1.8)&70.04 ($\pm$2.5)&20.28 ($\pm$7.6)&59.25 ($\pm$2.0)&60.19 ($\pm$1.8) \\

\hline
CB~\cite{cui2019class_cb_loss} &88.22 ($\pm$1.5)&70.36 ($\pm$1.7)&18.04 ($\pm$9.8)&58.88 ($\pm$2.7)&59.88 ($\pm$2.5) \\

LDAM-DRW~\cite{cao2019learning_LDAM} &92.53 ($\pm$0.6)&69.4 ($\pm$1.5)&24.55 ($\pm$9.1)&\underline{62.16 ($\pm$2.5)}&\underline{62.79 ($\pm$2.2)} \\

BSM~\cite{Ren2020balms} &91.4 ($\pm$0.7)&65.96 ($\pm$3.0)&26.54 ($\pm$7.7)&61.3 ($\pm$1.9)&61.7 ($\pm$1.6) \\

\hline
MixUp~\cite{zhang2018mixup} &94.23 ($\pm$0.6)&75.08 ($\pm$1.2)&3.93 ($\pm$3.3)&57.75 ($\pm$1.0)&59.25 ($\pm$0.9) \\

BalMixUp~\cite{10.1007/978-3-030-87240-3_31_balanced_mixup} &92.16 ($\pm$1.1)&74.57 ($\pm$1.7)&8.44 ($\pm$3.8)&58.39 ($\pm$1.1)&59.8 ($\pm$0.9) \\

\hline

BKD~\cite{ZHANG202336_bkd}  &92.53 ($\pm$0.9)&69.88 ($\pm$5.0)&17.43 ($\pm$12.6)&59.95 ($\pm$2.7)&60.81 ($\pm$2.3) \\

    \hline
\textbf{FoPro-KD (ours)}&92.78 ($\pm$2.0)&68.08 ($\pm$6.5)&31.9 ($\pm$8.5)&\textbf{64.25 ($\pm$0.8)}&\textbf{64.59 ($\pm$0.9)}
 \\
    \bottomrule
  \end{tabular}
      }
      \vspace{-0.5em}
\end{table}

%% file: tmi-5-Ablation.tex
\subsection{Ablation Studies}
\label{sec:abla}

\input{tmi-5.1-Main_Ablation}
\input{tmi-5.2-Exploitation_Effectiveness-Ablation}

\input{tmi-5.3-Exploration_Effectiveness}

\input{tmi-5.5-SmallerModels}

\input{tmi-5.7-FPG_Varying_ablation}

%% file: tmi-5.1-Main_Ablation.tex
\noindent\textbf{Effectiveness of EKD and FPG}
In~\Cref{tbl:isic_main_ablation}, we present an ablation study of our proposed components over the ISIC-LT. Our approach combines a Fourier prompt generator (FPG) with effective knowledge distillation (EKD) to exploit the \majorrevise{``free lunch model''}. Our experimental results on the ISIC-2019 dataset demonstrate that EKD alone improves performance by 1.5\%, 3.6\%, and 2.2\% on the ``MCC'' for the three imbalance ratios, respectively. By adding FPG, we achieve even higher performance gains of 4.5\%, 5.6\%, and 3.7\% on the ``MCC'' for class imbalance ratios of 1:100, 1:200, and 1:500 compared to the baseline, BSM~\cite{Ren2020balms}.


\begin{table}[htbp!]
\centering
\caption{Ablation of FLKD and FPG on three imbalance ratios on ISIC-LT}
\resizebox{1\columnwidth}{!}{
\begin{tabular}{l|c|c|ccc}
\toprule
\multirow{2}{*}{} &\multirow{2}{*}{EKD}& \multirow{2}{*}{FPG} & \multicolumn{3}{c}{Metric}    \\  
\cline{4-6}& & & MCC&Acc & F1-Score
   \\  \cline{4-6}
   & & & \multicolumn{3}{c}{ISIC-LT (1:100)}
    \\
   \hline
   BSM~\cite{Ren2020balms}& $\times$&$\times$ & 63.88 ($\pm$1.9) &68.15 ($\pm$1.7) &69.25 ($\pm$1.6)
 \\

w/ \textbf{EKD (ours)}& \checkmark&$\times$ 
&65.36 ($\pm$3.3) &69.47 ($\pm$2.9) &70.42 ($\pm$2.9)\\

\textbf{FoPro-KD (Ours)} &\checkmark& \checkmark& 
\textbf{68.33 ($\pm$2.3)} &\textbf{71.8 ($\pm$2.0)} &\textbf{73.88 ($\pm$1.9) } \\

\hline
   & & & \multicolumn{3}{c}{ISIC-LT (1:200)}
   \\
   \hline
      BSM~\cite{Ren2020balms}& $\times$&$\times$ & 60.47 ($\pm$1.6) &65.12 ($\pm$1.4) &66.2 ($\pm$1.2)  \\

w/ \textbf{EKD (ours)}& \checkmark&$\times$&
64.08 ($\pm$1.4) &68.35 ($\pm$1.2) &69.19 ($\pm$1.3)  \\

\textbf{FoPro-KD (Ours)} &\checkmark& \checkmark& 
\textbf{66.08 ($\pm$1.5)} &\textbf{69.8 ($\pm$1.3) }&\textbf{71.91 ($\pm$1.2)}\\

\hline
   & & & \multicolumn{3}{c}{ISIC-LT (1:500)}
   \\
   \hline
      BSM~\cite{Ren2020balms}& $\times$&$\times$ &53.61 ($\pm$1.1) &59.02 ($\pm$1.0) &60.27 ($\pm$0.9)\\

w/ \textbf{EKD (ours)}& \checkmark&$\times$&
55.81 ($\pm$1.6) &60.92 ($\pm$1.4) &62.02 ($\pm$1.3) \\

\textbf{FoPro-KD (Ours)} &\checkmark& \checkmark& 
\textbf{57.33 ($\pm$1.5)}&\textbf{61.9 ($\pm$1.5)}&\textbf{64.43 ($\pm$1.3)}\\
\bottomrule
\end{tabular}
}
\label{tbl:isic_main_ablation}
\vspace{-0.5em}
\end{table}






Our proposed EKD and FPG methods provide complementary benefits for improving the performance of the target model in the long-tailed setting. While FPG helps to explore the \majorrevise{``free lunch model''} latent space by explicitly asking what frequency patterns it wants in the input, EKD helps to exploit the \majorrevise{``free lunch model''} generalizable representation. By leveraging \majorrevise{``free lunch model''} frequency patterns, our approach achieves the best performance on the ISIC-LT dataset and HyperKvasir dataset, highlighting the importance of utilizing \majorrevise{``free lunch models''} for medical image classification with long-tailed class distributions.


%% file: tmi-5.2-Exploitation_Effectiveness-Ablation.tex


\noindent \textbf{Ablation of Free Factor}
We present an ablation study of the weighting factor, $\lambda_{f}$, for the exploitation proposed in~\Cref{eq:loss_total}, with experiments conducted on the ISIC-LT imbalance factor 1:500 without our proposed FPG. The results are summarized in~\Cref{tbl:free_ablation}.

\begin{table}[htbp!]
\centering
\caption{Exploitation $\lambda_{f}$ ablation without FPG on the ISIC-LT (Acc)}
    \resizebox{0.9\columnwidth}{!}{
 \begin{tabular}{l|cccc}
        \toprule
    {\multirow{2}{*}{Method}} &\multicolumn{4}{c}{ISIC-LT (1:500)}  
    \\
    \cline{2-5}
        & $\lambda_{f}=0$&$\lambda_{f}=1$ & $\lambda_{f}=3$ & $\lambda_{f}=5$ \\
        \toprule
        EKD & 
       59.02 ($\pm$1.0)&59.52 ($\pm$2.4) &\textbf{60.92 ($\pm$1.4)}&60.47 ($\pm$1.6) \\
        \bottomrule
    \end{tabular}
    \label{tbl:free_ablation}
    }
    \vspace{-0.5em}
\end{table}

We find that using effective knowledge distillation (EKD) with a factor of $\lambda_{f}=3$ improves the performance on the ISIC-LT dataset compared to the baseline ($\lambda_{f}=0$),~\cite{Ren2020balms}, achieving an ``Acc'' gain of 1.9\%. However, a higher value of $\lambda_{f}$ can deviate from the learning objective.

%% file: tmi-5.3-Exploration_Effectiveness.tex
\begin{table}[htbp!]
\centering
\caption{Ablation of Exploration on the ISIC-LT 1:500 dataset. VPT refers to spatial visual prompt tuning.}
\resizebox{1\columnwidth}{!}{
\begin{tabular}{l|c|c|ccc}

\toprule
\multirow{2}{*}{} & \multirow{2}{*}{$\mathcal{L}_{inv}$}& \multirow{2}{*}{$\mathcal{L}_{adv}$}&\multicolumn{3}{c}{Metric}\\  
\cline{4-6}& &&MCC & Acc &F1
   \\ 
   \hline
   EKD (ours)& 
$\times$&$\times$&  55.81 ($\pm$1.6) &60.92 ($\pm$1.4) &62.02 ($\pm$1.3)  \\

EKD + VPT& 
$\times$&$\times$&54.71 ($\pm$2.3)&  59.59 ($\pm$2.0)  &62.16 ($\pm$2.0) \\

EKD + FPG & 
\checkmark&$\times$ &
\underline{56.80 ($\pm$1.4)} &\underline{61.59 ($\pm$1.2)} &\underline{63.73 ($\pm$1.8)} \\
FoPro-KD& 
\checkmark&\checkmark&
\textbf{57.33 ($\pm$1.5)}&\textbf{61.9 ($\pm$1.5)}&\textbf{64.43 ($\pm$1.3)}\\

\bottomrule
\end{tabular}
}
\label{tbl:explore_ablation}
\vspace{-0.5em}
\end{table}

\noindent \textbf{Effectiveness of FPG} 
\majorrevise{To evaluate the importance of exploration with our proposed FPG, we perform an ablation study and report our results in~\Cref{tbl:explore_ablation}. First, we compare the results of our baseline, EKD, when employing conventional spatial visual prompting tuning (VPT) first introduced in~\cite{jia2022vpt} and adapted in closed-set source-free unsupervised domain adaptation~\cite{hu2022prosfda} through minimization of the batch normalization regularization technique detailed in~\Cref{eq:bn_reg}. Our findings indicate that using VPT with EKD decreases the performance compared to utilizing our proposed EKD alone. It is worth noting that establishing a meaningful open-set spatial domain mapping between the target medical dataset and the pre-training dataset, ImageNet, is a complex task. Furthermore, using a single prompt to learn this mapping leads to substantial alterations in the target dataset semantics and introduces training instability.} On the other hand, learning the FPG and exploring the \majorrevise{``free lunch model''} with only $\mathcal{L}_{inv}$ leads to an improvement over our proposed EKD with an increase of 1.0 \% and 1.7\% on ``MCC'' and  f1-score, respectively. Moreover, when using iterative adversarial knowledge distillation (AKD) along with $\mathcal{L}_{inv}$, we achieve the best performance with a notable gain of 1.5\%, 1.0\%, 2.4\% on the ``MCC'', ``Acc'', and F1 score respectively, compared to our proposed EKD. While $\mathcal{L}_{inv}$ ensures that the synthesizable Fourier amplitudes are representative of what the free lunch model wants, capturing the frequency patterns in the frequency bands it was trained on, $\mathcal{L}_{adv}$ is responsible for further exploring the latent space of the frozen model and making the frequency prompts more diverse than the ones previously distilled to the target model. 

\begin{figure}[htbp!]
\centering
\subfigure[]{
\includegraphics[width=.24\textwidth]{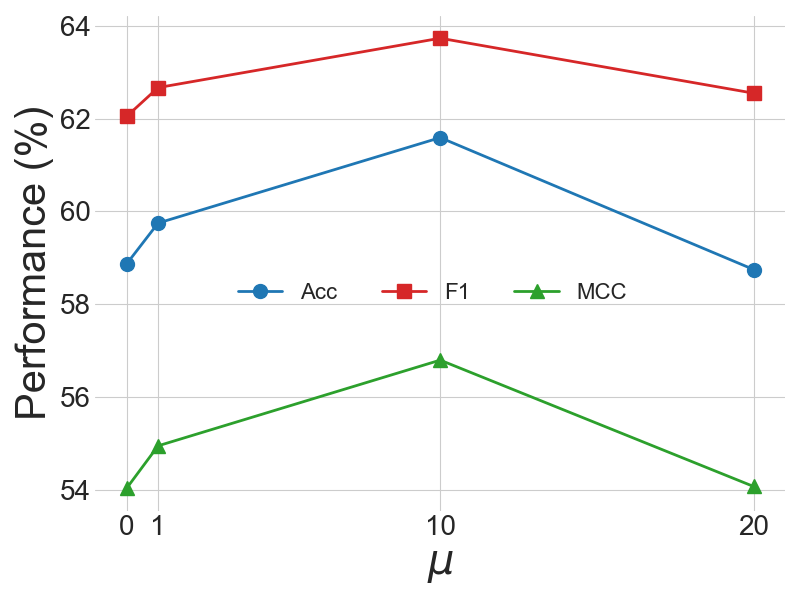}
}
\hspace{-2em}
\subfigure[]{
\includegraphics[width=.24\textwidth]{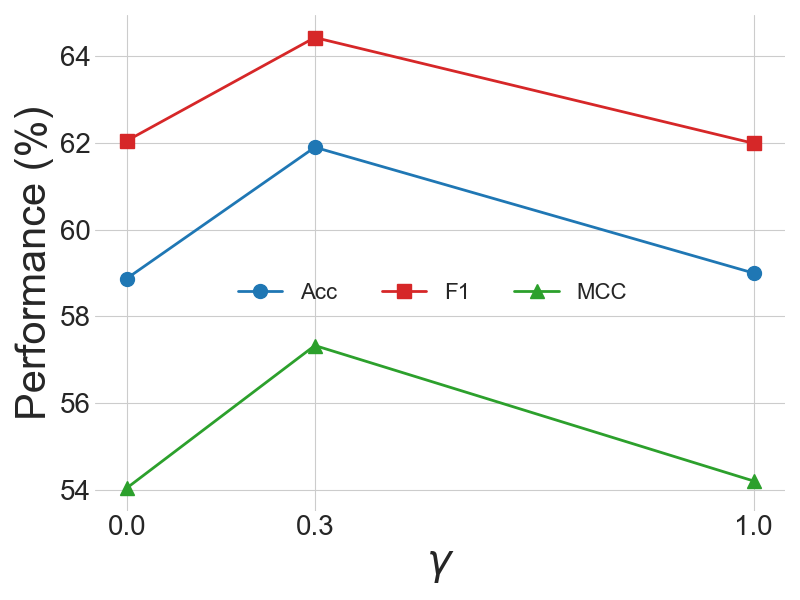}}
    \caption{Sensitivity to $\mu$ and $\gamma$ on the ISIC-LT Imbalance Ratio 1:500}
\label{fig:exploration_sensitivity}
\vspace{-0.5em}
\end{figure}

\noindent \textbf{Sensitivity of Balancing Regularization}
Batch normalization (BN) statistics are necessary for learning the Fourier prompts (FPG) in our proposed method. Similar to deep inversion and data-free knowledge distillation approaches~\cite{fang2021contrastive_cmi}, without BN, the FPG can be limited to balancing regularization. we perform ablation experiments on the balancing regularization weighting factor $\mu$ for the exploration phase proposed in ~\Cref{eq:inversion_loss} over the extremest ISIC-LT setting (1:500). As shown in~\Cref{fig:exploration_sensitivity} (a), we observe that a value of $\mu=10$ increases the performance by 2.6\%, 2.4\%, and 2.5\% on the ``MCC'', ``Acc'', and F1, respectively. Without using $\mu$, the exploration phase is limited to the BN statistics without activation of the free-lunch model latent space, which can limit the representation transfer. A high value of $\mu$, however, can negatively impact performance by encouraging the network to output a uniform distribution that is not discriminative nor informative.



\noindent \textbf{Sensitivity of AKD}
Next, we investigate the effect of the adversarial factor $\gamma$ proposed in~\Cref{eq:adv_loss} on the performance of the extremest ISIC-LT setting (1:500). We found that a low value of $\gamma$ (e.g., $\gamma=0.3$) can enhance performance by making the Fourier prompts more diverse with iterative adversarial training, increasing the performance by 1.0\% 2.0\% on the F1-score and ``MCC'', as shown in~\Cref{fig:exploration_sensitivity} (b). On the other side, a high value of $\gamma$ (e.g., $\gamma=1$) results in a 2.0\% drop in the F1-score. It is worth noting that, unlike other adversarial training approaches in domain adaptation, our focus is not on adversarial training but on synthesizing images based on the \majorrevise{``free lunch model''} by our proposed FPG. A lower value of $\gamma$ ensures the diversity of generated prompts, whereas a higher value may result in FPG generating worst-case images with random amplitudes that the \majorrevise{``free lunch model''} cannot comprehend, leading to a decrease in overall performance.

%% file: tmi-5.5-SmallerModels.tex
\begin{figure}[htbp!]
    \centering
    \includegraphics[width=\columnwidth]{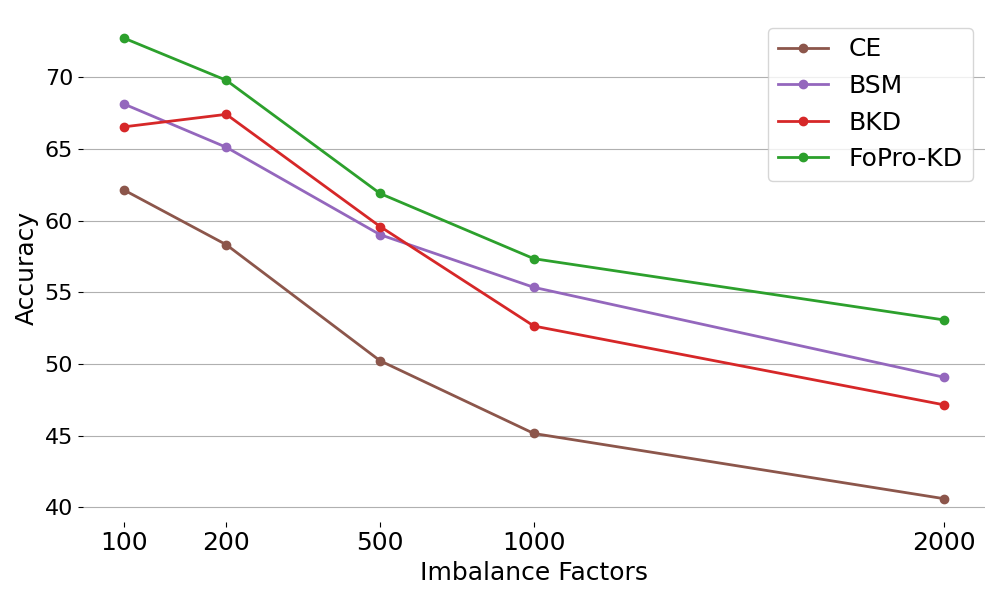}
    \caption{\majorrevise{Effectiveness of FoPro-KD across Imbalance Factors, including 1:1000 and 1:2000, on the accuracy ``Acc'' metric for ISIC-LT dataset.}}
    \label{fig:ablation_imbalance_factors}
\end{figure}

\majorrevise{
\noindent \textbf{Effectiveness of Addressing Extreme Class Imbalance} We assess the impact of severe class imbalance on the performance of our method, FoPro-KD. Specifically, we evaluate the accuracy (``Acc'') for extreme imbalance factors of 1:1000 and 1:2000 using the ISIC-LT balanced test-set and present the results in~\Cref{fig:ablation_imbalance_factors}. Notably, as the severity of class imbalance increases, our method, FoPro-KD, consistently outperforms the state-of-the-art curriculum-based method, BKD~\cite{ZHANG202336_bkd}, demonstrating remarkable improvements of 4.7\% and 5.9\% on the ``Acc'' metric for imbalance factors of 1:1000 and 1:2000, respectively.}

\majorrevise{
Furthermore, we observe in~\Cref{fig:ablation_imbalance_factors} that the baseline, BSM~\cite{Ren2020balms}, outperforms BKD~\cite{ZHANG202336_bkd}, by 2.7\% and 1.9\% over the imbalance factors of 1:1000 and 1:2000 respectively. In extremely imbalanced scenarios, BKD~\cite{ZHANG202336_bkd} as a curriculum-based method encounters challenges stemming from the inherent bias in representation learning. In such cases, the features learned in the initial stages tend to be heavily influenced by the majority class, exacerbating the imbalance issue for the minority classes~\cite{bai2023on_hualiang_iclr23}. In contrast, our approach introduces a novel perspective on cross-task knowledge compression from a publicly available pre-trained model on natural images, which does not suffer from this inherent bias. Moreover, it is worth noting that BKD~\cite{ZHANG202336_bkd} operates in the logit space instead of the feature space, which introduces additional challenges related to classifier bias towards the majority class~\cite{Kang2020Decoupling}. By operating in the feature space with a cross-task pre-trained network, our method mitigates these bias-related challenges, enabling more effective handling of extreme class imbalance.}

\par

\noindent \textbf{EKD benefits LT even with smaller models} The learning of the target model can be limited with an upper bound to the capacity of the free lunch model, $f$, and the MLP projector, and the information gained from $f$ to the target task. However, we demonstrate in~\Cref{tbl:abl:ekd_small} that such limitations do not adversely affect the performance of the minority class on ISIC-19 LT (``Tail''), with linear probed (LP) supervised ImageNet weights achieving 41.89\% and EKD achieving 55.93\% on the ``Tail'' accuracy. 

\begin{table}[htbp!]
\footnotesize
\centering
\caption{Effective Knowledge Distillation (EKD) with varying free lunch models. Results are averaged across 5 runs and across the three imbalance ratios (1:100, 1:200, 1:500) on the ISIC-LT dataset. \majorrevise{LP+FPG denotes linear probing on the optimized FPG+$f$}.}
\vspace{-0.5em}
\label{tbl:abl:ekd_small}
\resizebox{1\columnwidth}{!}{
\begin{tabular}{lcccc|cc}
\toprule
\multicolumn{5}{c}{Setting}
&\multicolumn{2}{c}{Metric (\%)}
\\
\cline{1-5} \cline{5-7}
Method
&Target
&Target Init &Free Lunch& Free Lunch Init&Tail&MCC
\\
\hline
LP&None&None&RN-50&Sup-ImageNet& 41.89&48.77 \\
LP&None&None&RN-50& MoCov2&48.72&49.85
\\
LP+FPG&None&None&RN-50& MoCov2&48.46&51.34
\\
\hline
FT&None&None&RN-50&MoCov2 &27.6&56.0 \\
FT&None&None&RN-50&Sup-ImageNet&35.93&62.51
\\
\hline 
BSM &RN-18 &None&None&None&46.6&59.32\\
EKD&RN-18&None&RN-50&Sup-ImageNet&51.47&61.13\\
\hline
BSM&RN-18 &ImageNet&None&None&52.07&66.24\\

EKD&RN-18&ImageNet&\underline{RN-18}&Sup ImageNet&\underline{55.87}&\underline{67.18}\\

EKD&RN-18&ImageNet&\textbf{RN-50}&Sup ImageNet&\textbf{55.93}&\textbf{68.09}\\

\bottomrule
\end{tabular}
}
\vspace{-1em}
\end{table} 

Our experiments presented in~\Cref{tbl:abl:ekd_small} demonstrate that our proposed EKD method can improve the performance of the target task even with smaller models. Specifically, we show that when given a target model $g$ and its pre-trained version as the free lunch model $f$, EKD can benefit the tail classes using the frozen features from $f$ despite $f$ having the same capacity as $g$ and being pre-trained on ImageNet. We observed a performance gain of 3.8\% on tail class accuracy and 1.0\% on ``MCC'' compared to the best-performing baseline initialized with ImageNet weights. It is worth mentioning that these results are averaged over 5 runs over the 3 class imbalance ratios (15 experiments). This phenomenon arises because fine-tuning can distort the pre-trained features, leading to a drop in generalization performance. However, the target model can further enhance its performance by using free discriminative distribution during training. While EKD can improve the performance of the target task even with smaller models, the best performance is achieved when using ResNet50 (RN50) as the free lunch model ($f$), with a performance gain of 1.85\% and 1.81\% on ``MCC'' compared to the baseline, BSM~\cite{Ren2020balms}, when the target model is initialized randomly (None) or with ImageNet weights respectively. While most empirical evaluations ignore pre-trained initialization to provide fair and better convergence analysis, initialization unsurprisingly increases the averaged performance by 6.92\% and 6.96\% on ``MCC'' for the baseline and our proposed EKD, respectively. 

\majorrevise{Furthermore, we present the classification performance results achieved through linear probing of the pre-trained model using our proposed Fourier Prompt Generator (FPG). The FPG demonstrates a significant enhancement of 1.49\% in the Matthews Correlation Coefficient (MCC), which is particularly sensitive to class imbalance when compared to the baseline model (LP-MoCoV2). This improvement underscores the effectiveness of the FPG in enhancing the overall feature representation of the free lunch model. However, we do observe a slight performance decline of 0.26\% specifically in the tail class. This decrease can be attributed to the modifications in the styles present within the tail class, dermatofibroma (DF). DF benefits from color information, exhibiting a range from pink to light brown in lighter skin tones and from dark brown to black in darker skin tones. Importantly, this style alteration does not affect the knowledge distillation process, as irrelevant styles and biases are exclusively imposed on the projector component (MLP) of the target network in the projected feature space of the free lunch model within the knowledge distillation framework, thereby not impacting the classifier (FC) of the target network.
}

%% file: tmi-5.7-FPG_Varying_ablation.tex
\begin{figure}[htbp!]
    \centering
\includegraphics[width=\columnwidth]{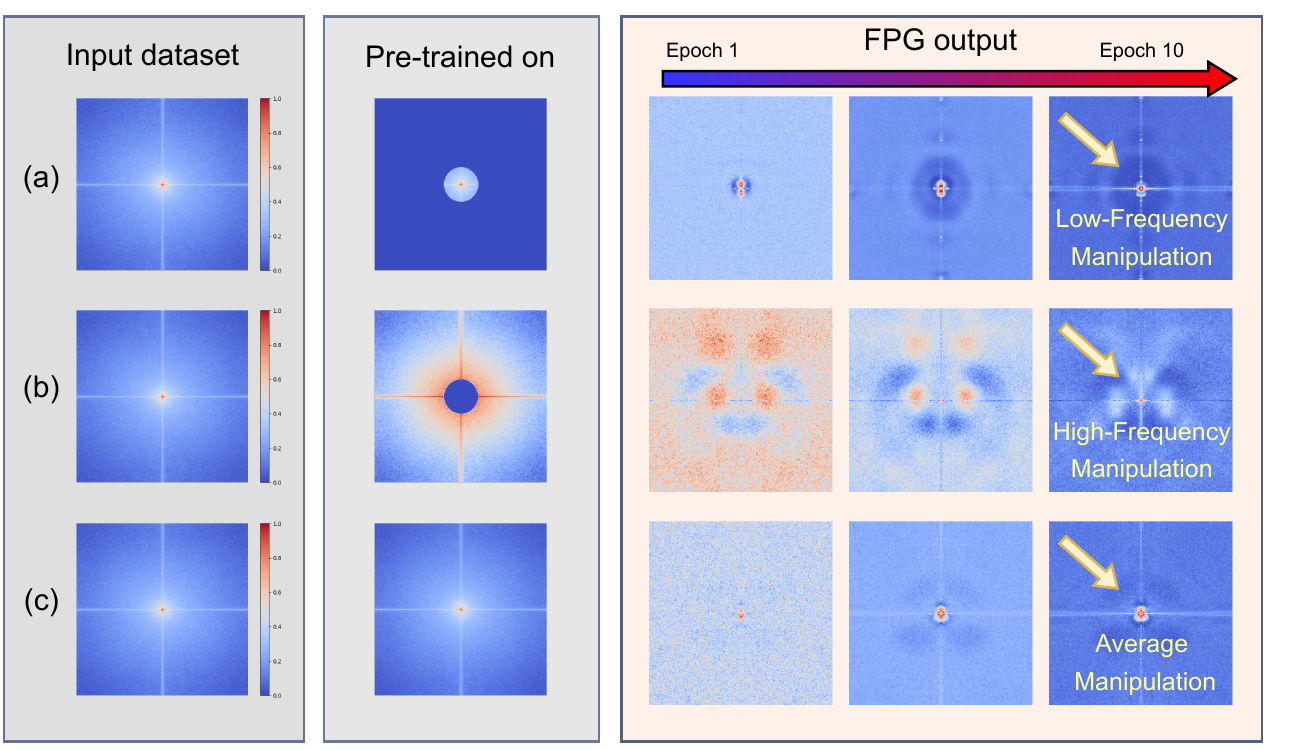}
    \caption{Average FPG generated prompts in three scenarios of pre-training $f$ on different frequency components of the ISIC-LT dataset. (a) Pre-training $f$ on only the low-frequency components. (b) Pre-training $f$ on only the high-frequency components. (c) Pre-training $f$ on all-frequency components.}
    \label{fig:abl:FPG_big_plot}
\end{figure}

\noindent \textbf{FPG is conditional on both the input and the pre-trained model} 
    In~\Cref{fig:abl:FPG_big_plot}, we demonstrate the behavior of our FPG with only $\mathcal{L}_{explore}$~\Cref{eq:inversion_loss}. In (a), we trained the frozen pre-trained model, $f$, on only the low-frequency components of the ISIC-LT dataset. We observed that the FPG converged to a similar average amplitude as the input dataset but with different \majorrevise{enhancement} in the low-frequency parts that are conditional on $f$. (b) shows the FPG's behavior when $f$ was trained on only the high-frequency components of the ISIC-LT dataset. We found that the FPG attends to the different frequencies in their higher frequencies that $f$ has captured. Finally, in (c), we trained $f$ on all frequency components of the input dataset. Interestingly, we found that the average amplitude generated by the FPG does not fully reduce to the amplitude of the source dataset, although it is conditional on the input dataset. This is because we do not have any prior knowledge of what frequency patterns in what frequency bands the \majorrevise{``free lunch model''} extracts from the dataset in the pre-training stage. Nonetheless, FPG was able to amplify or suppress certain frequencies to provide understanding and interpretation of the behavior of pre-training models.

Our proposed Fourier Prompts Generator (FPG) is designed for understanding and interpreting the behavior of \majorrevise{``free lunch models''}. Unlike prior methods that rely on adding noise to synthesize worst-case images, the FPG is conditional on both the input dataset and the \majorrevise{``free lunch model''}. 
Our method leverages the different frequency patterns captured in the pre-training stage of the network to amplify or suppress certain frequencies. By exploring these patterns with the FPG, we can gain insights into the specific input preferences of the \majorrevise{``free lunch model''} that enable better representational transfer and interoperability.

%% file: tmi-6-DiscussionAndConc.tex
\section{Discussion}

Rare disease classification is a crucial aspect of medical imaging, and leveraging publicly available pre-trained models can potentially improve the diagnosis and representations of these diseases. Existing work in this area often regularizes training on synthesizing worst-case scenarios and extracting the knowledge using closed-set datasets, without fully exploiting the generalization capabilities of widely known pre-trained models. Although some studies have explored effective prompting techniques for these models, their approaches are often limited to high-level features and prompt engineering without a deep understanding of how these ``free lunch'' encoders work, or how their representations can be further enhanced through a fundamental understanding of DNNs. In this work, we address this gap by investigating an intuitive phenomenon that has been widely neglected in the community: explicitly asking the pre-trained model what it wants, conditional on a cross-task medical input data, in order to gain insights into the learning dynamics of these models for effective representation learning. Through our method, we successfully demonstrate and leverage this phenomenon, shedding light on the inner workings of these models' frequency patterns and their behavior toward representation learning. \majorrevise{

In our work, we have discovered that learning frequency prompts facilitate knowledge distillation. This motivates future work on prompting large vision models on a frequency basis for several downstream tasks. Furthermore, we find that ramping up significantly adversarial training can have a negative impact on the training process rather than enhancing it. Our primary goal is to enhance the frequency of image inputs by prompting on a frequency basis. Thus challenging the network with adversarial training impacts the performance. Additionally, we find that EKD benefits the medical imaging long-tailed problem even when using smaller ``free lunch models''. This can be attributed to the fact that fine-tuning can distort pre-trained features, leading to a decrease in generalization performance~\cite{kumar2022finetuning_LP_FT}.

Our FoPro-KD demonstrates a higher percentage increase compared to all other methods when decreasing the imbalance ratio. For instance, even with a significant class imbalance of 1:2000, the performance of the model remains considerably higher than random accuracy. It is important to note that if the imbalance factor is pushed to values smaller than 1:12725, it indicates the network's inability to recognize samples it has not encountered before, reducing the problem to out-of-distribution detection.
 }
 
Our proposed approach has certain limitations. One such limitation is the need for a better understanding of the learning dynamics of deep neural networks (DNNs) and the conditions required for capturing these frequency patterns on more sophisticated models~\cite{Makino2020DifferencesBH}. 
Additionally, while skin lesions and gastrointestinal images can be considered out-of-distribution data for the free lunch model, there are extreme cases in medical imaging, such as X-rays and MRIs, which may require further exploration. Future research should aim to bridge the gap between natural image and medical imaging domains to enhance our understanding of the billions of parameters utilized in pre-trained models released every year.



\section{Conclusion}
In conclusion, our proposed FoPro-KD method provides an effective and efficient approach for compressing knowledge from publicly available pre-trained models to medical image classification tasks. 
We believe that future research should continue to explore the generalization capabilities of these largely available pre-trained models and develop methods to compress their knowledge for medical imaging tasks while preserving their generalization capabilities. Our method's ability to utilize the pre-trained model's knowledge to smaller target models for medical tasks can be particularly useful in clinical settings where affordable AI is needed. Overall, we believe that our FoPro-KD method represents a promising direction for addressing long-tailed classification problems and transfer learning in medical imaging.